\documentclass[%
 aip,
jcp,% short, numerical bibliography, 
floatfix,
 amsmath,amssymb,
 reprint,%
 groupedaddress,
]{revtex4-1}

\usepackage{subfigure}
\usepackage{graphicx}% Include figure files
\usepackage{braket}
\usepackage{dcolumn}% Align table columns on decimal point
\usepackage{bm}% bold math
\begin{document}

\preprint{AIP/123-QED}

%\title[Sample title]{Sample Title:\\with Forced Linebreak\footnote{Error!}}% Force line breaks with \\
%\thanks{Footnote to title of article.}

%\author{A. Author}
% \altaffiliation[Also at ]{Physics Department, XYZ University.}%Lines break automatically or can be forced
%with \\
%\author{B. Author}%
% \email{Second.Author@institution.edu.}
%\affiliation{ 
%Authors' institution and/or address%\\This line break forced with \textbackslash\textbackslash
%}%

% \homepage{http://www.Second.institution.edu/~Charlie.Author.}
%\affiliation{%
%Second institution and/or address%\\This line break forced% with \\
%}%

\title{Nonadiabatic Ehrenfest molecular dynamics within the projector augmented-wave method}

%\author{Ari Ojanper\"a}
%\author{Ville Havu}

%\affiliation{COMP/Applied Physics, Aalto University, P.O. Box 11100, FI-00076 AALTO, Finland}
\author{Ari Ojanper\"a,$^1$ Ville Havu,$^1$ Lauri Lehtovaara,$^2$ and Martti Puska$^1$}
\affiliation{$^1$COMP/Applied Physics, Aalto University, P.O. Box 11100, FI-00076 AALTO, Finland}
\affiliation{$^2$LPMCN, Universit\'e Claude Bernard Lyon I, F69622 Villeurbanne Cedex, France}
% \author{Lauri Lehtovaara}
% %\email[]{ari.ojanpera@aalto.fi}
% \affiliation{LPMCN, Universit\'e Claude Bernard Lyon I, F69622 Villeurbanne Cedex, France }
% 
% \author{Martti Puska}
% 
% \affiliation{COMP/Applied Physics, Aalto University, P.O. Box 11100, FI-00076 AALTO, Finland}

%\affiliation{COMP/Applied Physics, Aalto University, P.O. Box 11100, FI-00076 AALTO, Finland}


%\thanks{}


\date{\today}% It is always \today, today,

\begin{abstract}
We have derived equations for nonadiabatic Ehrenfest molecular dynamics which conserve the total energy in the
case of time-dependent discretization for electrons. A discretization is time-dependent in all cases where it
or part of it depends on the positions of the nuclei, for example, in atomic orbital basis sets, and in the
projector augmented-wave (PAW) method, where the augmentation functions depend on the nuclear positions. We
have derived, implemented, and analyzed the energy conserving equations and their most common approximations
for a 1D test system where we can achieve numerical results converged to a high accuracy. Based on the
observations in 1D, we implement and analyze the Ehrenfest molecular dynamics in 3D using the PAW method and
the time-dependent density functional formalism. We demonstrate the applicability of our method by carrying
out calculations for small and medium sized molecules in both the adiabatic and the nonadiabatic regime.

\end{abstract}

%\pacs{02.70.Ns, 31.15.-p, 31.15.ee, 31.15.xg, 71.15.Ap}% PACS, the Physics and Astronomy
%\keywords{Nonadiabatic molecular dynamics, Ehrenfest molecular dynamics, PAW, TDDFT}%Use showkeys class
%option
%if keyword
\maketitle



\section{Introduction}

Many natural processes, such as light absorption, ignition of chemical reactions and ion-atom collisions, are
related to excited electronic states and their time development. The most general approach for treating such
nonadiabatic processes, which in general involve two or more coupled electronic states, would be to solve the
time-dependent many-body Schr{\"o}dinger equation. However, this is not feasible for systems consisting of
more than a few electrons. Moreover, the standard ab initio mole\-cular dynamics (AIMD) methods such as
Car-Parrinello MD \cite{Car1985} (CPMD)  or Born-Oppenheimer MD (BOMD) (see, for example, Ref.
\onlinecite{Marx2009}) confine electrons to a single adiabatic state, typically the ground state.
Semiclassical methods, such as Ehrenfest molecular dynamics (Ehrenfest MD) or trajectory surface hopping
(TSH), in which the electrons are treated quantum-mechanically via the time-dependent Schr{\"o}dinger equation
and the nuclei classically via the Newtonian mechanics, have been deve\-loped some decades ago, but only
during the last decade they have become feasible in atomistic simulations beyond small systems. This is partly
due to the methodo\-logical advances in time-dependent density functional theory \cite{Yabana1996} (TDDFT)
which provides a computationally affordable basis for the Ehrenfest MD and the TSH methods, and also due to
the rapidly increasing amount of computational resources available. 

Ehrenfest MD within TDDFT offers a simple yet effective framework for simulating nonadiabatic processes by
coupling the time-dependent Kohn-Sham (TDKS) equations \cite{Runge1984} with classical equations of motion
for nuclei via the KS potential energy surface (PES). The method works well for condensed matter, where many
single
electron levels are involved and a single reaction path dominates the nonadiabatic process such as in carbon
nano\-structures. However, when reactions pass regions of close lying electronic states but end up in a state
which is well described by a single potential energy surface, the TSH method is preferable
\cite{Tapavicza2007}. This is due to the deficiency of the Ehrenfest MD that the system remains in a mixed
state after exiting the nonadiabatic region. Moreover, due to its mean-field character, the Ehrenfest MD
cannot correctly describe multiple reaction paths \cite{Tully1998}.

The Ehrenfest MD has been succesfully used for studying various nonadiabatic processes such as collisions
bet\-ween atomic oxygen and graphite clusters \cite{Isborn2007}, excited carrier dynamics in carbon nanotubes
\cite{Miyamoto2006} and electronic excitations in ion bombardment of carbon nanostructures
\cite{Krasheninnikov2007}. Previously, there have been only a few implementations of the Ehrenfest MD, all of
which are based on pseudopotentials. Various basis sets such as LCAO \cite{Meng2008}, plane waves
\cite{Sugino1999} and real space grids \cite{Andrade2009} have been used. Real space techniques have several
advantages: 1) the qua\-lity of the results can be controlled by a single variable, the grid spacing, 2)
parallelization can be done straightforwardly using domain decomposition and 3) different boundary conditions
such as Dirichlet, periodic, or a mixture of them can be easily applied. 

In spite of the extensive use of the the projector augmented-wave \cite{Blochl1994} method in
electronic structure calculations, to our knowledge, it has not been used in Ehrenfest MD simulations
previously. Compared to pseudopotentials, the PAW method improves the description of the transition metal
elements and the first row elements with open $p$-shells. Moreover, in order to perform accurate calculations,
the PAW method in real space allows one to use fewer grid points than pseudopotentials \cite{Walter2008} as
well as longer time steps in the propagation of the TDKS equations \cite{Walker2007}. The all-electron (AE)
nature of the PAW method, albeit with the core states frozen, is also a methodological advantage over
pseudopotentials. However, compared to pseudopotentials, Ehrenfest MD within the PAW method is more
complicated due to the augmentation functions that depend on the atomic positions. First, an additional
term describing the mo\-ving spatial gauge of the electrons must be included in the TDKS equations
\cite{Qian2006}. Also, the Hellmann-Feynman (HF) theorem (see, for example, Ref. \onlinecite{Perdew2003}),
traditionally used for calculating the atomic forces, is no longer valid.

The present paper is constructed as follows. In Sec. \ref{sec:methodology}, we investigate various
expressions for the Ehrenfest MD forces and derive results regarding the total energy conservation in the
quantum-classical dynamics. Moreover, we describe our propagation algorithm for the Ehrenfest MD equations. In
Sec. \ref{sec:efmd_fem}, we present the one-dimensional test system and the example calculations carried out
with it. The Ehrenfest MD implementation within the PAW method is described in Sec. \ref{sec:efmd_gpaw}. The
applicability of our method is demonstrated by carrying out simulations for the NaCl dimer and the H$_2$C =
NH$_2^+$ and C$_{40}$H$_{16}$ molecules. Furthermore, the applicability of the different forces in both the
adiabatic and the nonadiabatic regime is discussed. Finally, we give brief conclusive remarks in Sec.
\ref{sec:discussion}.

\section{Methodology} 
\label{sec:methodology}
%\label{}
Ehrenfest MD in general can be defined by the time-dependent Schr\"odinger equation for the electrons
and the Newton equations of motion for the nuclei (atomic units are used throughout this paper)
\begin{eqnarray}
 i \frac{\partial \Psi}{\partial t} =&& {\cal H}_{\text{el}} (\lbrace{\bf r}_i \rbrace, \lbrace{\bf
R}_a\rbrace)
\Psi (\lbrace{\bf r}_i \rbrace, t; \lbrace{\bf R}_a \rbrace), \label{eq:efmd_tdse}\\
M_a \ddot{{\bf R}}_a = -&& \nabla_{{\bf R}_a}  E_{\text{eff}} = -
\nabla_{{\bf{R}}_a} \braket{\Psi | {\cal H}_{\text{el}} |
\Psi},\label{eq:efmd_newton}
\end{eqnarray}
where $\Psi$ is the many-particle electronic wavefunction, depending explicitly on the time $t$ and
the electronic
degrees of freedom $\lbrace {\bf r}_i \rbrace$, and implicitly on the atomic positions $\lbrace {\bf R}_a
\rbrace$. ${\cal H}_{\text{el}}$ is the electronic Hamiltonian. Thus,
the force on the nuclei is calculated as an average over all electronic
adiabatic states, i.e., the Ehrenfest MD is effectively a mean-field theory. 
By expanding the electronic wavefunctions in the basis obtained by solving the
time-independent Schr\"odinger equation, one can show that nonadiabatic effects are included in the Ehrenfest
MD \cite{Marx2009}. In this section, we investigate the Ehrenfest MD within the single-particle formalism in a
finite basis.

\subsection{A general time-dependent quantum-classical system} \label{ssec:gen_td}

We consider a general time-dependent electronic system within Kohn Sham-like single-particle formalism when
the Hamiltonian operator $\hat{H}$ can be written as a sum of a position-dependent term $\hat{H}^0$
and a nonlinear term $\hat{H}'$ that contains terms depending on the electronic density $\rho$
\begin{equation}
 \hat{H} ({\bf r}, \lbrace{\bf R}_a\rbrace, \rho)= \hat{H}^0 ({\bf r}, \lbrace{\bf R}_a\rbrace) + \hat{H}'
({\bf
r}, \rho). \label{eq:h_gen}
\end{equation} 
The energy functional of the system is defined in terms of $\rho$ as
\begin{equation}
 E_{\text{el}} [\rho; \lbrace{\bf R}_a\rbrace] = T_s [\rho] + E_{\text{ext}} [\rho; \lbrace{\bf R}_a\rbrace]  
+ E'[\rho],\label{eq:eel}
\end{equation}
where $T_s$ is the kinetic energy of the non-interacting electrons, $E_{\text{ext}}$ is the energy due to the
external potential which we assume to depend explicitly on the atomic positions, and $E'$ is a functional that
contains nonlinear density-dependent terms. The electronic density $\rho$ is a function of atomic
positions $\lbrace{\bf R}_a\rbrace$ and time $t$. We use the semicolon in Eq. (\ref{eq:eel}) to
distinguish between the function and vector dependencies of $E_{\text{el}}$. In the Kohn-Sham formalism, $E'$
contains the Hartree and exchange-correlation energies. $\hat{H}'$ [Eq. (\ref{eq:h_gen})] is then the
functional derivative of $E'[\rho]$ with respect to the density. We expand the single-particle electronic
states on a basis $\lbrace \chi_k \rbrace$ that depends explicitly
on the atomic positions, 
\begin{equation}
 \psi_n ({\bf r}, \lbrace{\bf R}_a\rbrace, t)  = \sum_k c_{nk} (t) \chi_k ({\bf r}, \lbrace{\bf R}_a\rbrace).
\label{eq:psin}
\end{equation}
Moreover, we assume that the time-dependency of the basis functions is solely due to the movement of the
atomic positions. With this construction, we define the Hamiltonian matrices ${\bf H}$, ${\bf H}^0$
and ${\bf H}'$ as follows
\begin{eqnarray}
H_{ij} =&& \braket{\chi_i | \hat{H} | \chi_j}, \\
 H^0_{ij} =&& \braket{\chi_i | \hat{H}^0 | \chi_j}, \\
 H'_{ij} =&& \braket{\chi_i | \hat{H}' | \chi_j}. \label{eq:hprime}
\end{eqnarray}

We write the total Hamiltonian of the quantum-classical system consisting of quantum-mechanical electrons and
classical nuclei as
\begin{equation}
 \mathcal{H} = \sum_a \frac{{\bf p}_a^2}{2 M_a} + E_{\text{el}}.
\end{equation}
In quantum-classical molecular dynamics, it is essential that the time derivative of the total
Hamiltonian $\mathcal{H}$ is at least approximatively zero, i.e., $\mathcal{H}$ is an invariant. The time
derivative of the total Hamiltonian reads as
\begin{equation}
 \frac{d \mathcal{H}}{d t} = \sum_a {\bf v}_a \cdot {\bf F}_a + \frac{d E_{\text{el}}}{d
t}, \label{eq:detot_dt}
\end{equation}
where ${\bf v}_a$ is the velocity of atom $a$, and ${\bf F}_a$ is the atomic force.
Now, the time derivative of the electronic energy [Eq. (\ref{eq:eel})] can be computed as
\begin{equation}
\frac{d E_{\text{el}}}{dt} = \sum_a {\bf v}_a \cdot \frac{\partial E_{\text{el}}}{\partial {\bf R}_a} +
\frac{\partial E_{\text{el}}}{\partial t}.  \label{eq:deel_dt}
\end{equation}
In order to proceed from the above expression, we expand the electronic states $\psi_n$ in the time-dependent
single-particle Schr{\"o}dinger equation,
\begin{equation}
 i \frac{\partial \psi_n}{\partial t} = \hat{H} \psi_n, \label{eq:td_se}
\end{equation}
on the basis $\lbrace \chi_k \rbrace$ [Eq. (\ref{eq:psin})]. Consequently, we arrive at the following matrix
representation of the single-particle TD Schr{\"o}dinger equation [Eq. (\ref{eq:td_se})]
\begin{equation}
 i {\bf S} \frac{\partial {\bf c}_n}{\partial t} = ({\bf H} + {\bf P}) {\bf c}_n, \label{eq:tdse_basis}
\end{equation}
where the vector ${\bf c}_n$ contains the basis function coefficients of state $n$, $c_{nk}$,
and the matrices ${\bf S}$ and ${\bf P}$ are defined as
\begin{eqnarray}
 S_{ij} =&& \braket{\chi_i | \chi_j}, \\
 P_{ij} =&& -i\braket{\chi_i | \frac{\partial \chi_j}{\partial t}}.
\end{eqnarray}
The ${\bf S}$ matrix describes the overlap between the basis functions, while the ${\bf P}$ matrix
takes into account the moving spatial gauge of the electrons due to the changing atomic positions and
conserves the norm of the electronic states. It disappears if the nuclei do not move. Furthermore, employing
the chain rule, the ${\bf P}$ matrix can be written in terms of the atomic velocities as 
\begin{equation}
 {\bf P} = -i \sum_a {\bf v}_a \cdot {\bf  D}_a,
\end{equation}
with the definition
\begin{equation}
 D_{a,ij} = \braket{\chi_i | \frac{\partial \chi_j}{\partial {\bf R}_a}}.
\end{equation}
Using the matrix representation of the TD Schr{\"o}dinger equation [Eq. (\ref{eq:tdse_basis})], we can compute
the time derivatives of the vectors ${\bf c}_n$. Moreover, as the nonlinear energy term $E'$ [Eq.
(\ref{eq:eel})] is essentially of the form
\begin{equation}
 E' = \int F({\bf r}, \rho({\bf r})) d {\bf r},
\end{equation}
where $F$ is a density-dependent function. Using the chain rule, we can compute the (partial) time derivative
of the nonlinear energy,
\begin{equation}
 \frac{\partial E'}{\partial t} = \int \frac{\partial F}{\partial \rho}\frac{\partial \rho}{\partial t} d{\bf
r}.
\label{eq:depdt}
\end{equation}
Since $\partial F / \partial \rho$ equals the functional derivative of $E'$ with respect to the density, 
Eq. (\ref{eq:depdt}) can be written in the following matrix form
\begin{equation}
 \frac{\partial E'}{\partial t} = \sum_n [{\bf c}_n^{\ast} {\bf H'} \frac{\partial {\bf c}_n}{\partial t} +
c.c.]. \label{eq:depdt_v2}
\end{equation}
Similar reasoning can be applied to the linear part of the electronic energy, $E^0 = E_{\text{el}} - E'$.
Thus, we get the following expression for its time derivative
\begin{equation}
  \frac{\partial E^0}{\partial t} = \sum_n [{\bf c}_n^{\ast} {\bf H^0} \frac{\partial {\bf c}_n}{\partial t} +
c.c.]. \label{eq:de0dt}
\end{equation}
With Eqs. (\ref{eq:depdt_v2}) and (\ref{eq:de0dt}) and the matrix representation of the single-particle TD
Schr{\"o}dinger equation [Eq. (\ref{eq:tdse_basis})], the time derivative of the electronic energy [Eq.
(\ref{eq:deel_dt})] takes the following form
\begin{eqnarray}
&&\frac{d E_{\text{el}}}{dt} = \sum_a {\bf v}_a \cdot \frac{\partial
E_{\text{el}}}{\partial {\bf R}_a} + \frac{\partial E_{\text{el}}}{\partial t} \nonumber \\
=&& \sum_{a} {\bf v}_a \cdot\left[ \frac{\partial E_{\text{el}}}{\partial {\bf R}_a} -\sum_{n}  
{\bf c}_n^{\ast} ( {\bf H} {\bf S}^{-1} {\bf D}_a + c.c.) {\bf c}_n \right].
\label{eq:deel_dt_final}
\end{eqnarray}
The dynamics of the quantum-classical system is defined by the Newton equations for the nuclei
and Eq. (\ref{eq:tdse_basis}) for the electrons, i.e.,
\begin{eqnarray}
M_a \ddot{{\bf R}}_a =&& {\bf F}_a, \label{eq:newton_gen} \\
 i {\bf S} \frac{\partial {\bf c}_n}{\partial t} =&& ({\bf H} + {\bf P}) {\bf c}_n.
\label{eq:td_se_gen}
\end{eqnarray}
Now, whether the total Hamiltonian $\mathcal{H}$, usually interpreted as the total energy of the system,
is a conserved quantity or not, depends on the type of force ${\bf F}_a$ used in the calculations. Next,
we will consider three different forces: 1) Total energy conserving (EC) force, 2) Incomplete basis set
corrected (IBSC) force and 3) Hellmann-Feynman (HF) force.

\subsubsection{The total energy conserving force}
The most general Ehrenfest MD force expression can be derived using the requirement that the time derivative
of the total Hamiltonian [Eq. (\ref{eq:detot_dt})] equals zero. This can be achieved by defining the force by
the terms inside the brackets in Eq. (\ref{eq:deel_dt_final}) with an opposite sign, i.e.,
\begin{equation}
  {\bf F}_a^{\text{EC}} = -\frac{\partial E_{\text{el}}}{\partial {\bf R}_a}
+\sum_n  {\bf c}_n^{\ast}({\bf H} {\bf S}^{-1} {\bf D}_a + c.c.) {\bf c}_n.
\label{eq:f_econs}
\end{equation}
A very similar expression has been derived using first the HF theorem to find an equation for
the atomic forces according to the Ehrenfest MD, and then writing the resulting equations in a basis
\cite{Doltsinis2002}. Moreover, Di Ventra \cite{DiVentra2000} suggested that for time-dependent processes, one
should calculate the force acting on atom $a$ as the time derivative of the expectation value of its momentum
operator, i.e.,
\begin{equation}
  {\bf F}_a^{\text{TD}} = -i \frac{d}{dt} \sum_n  \braket{\psi_n | \frac{\partial}{\partial {\bf R}_a} |
\psi_n}. 
\label{eq:f_diventra}
\end{equation}
It can be shown that the EC force [Eq. (\ref{eq:f_econs})] follows from the TD force expression.
Next, we derive the HF and IBSC forces that are approximations to the EC force.

\subsubsection{The IBS corrected force} \label{ssec:ibsc}
By assuming that the wavefunctions are the eigenstates of the Hamiltonian operator, i.e., in matrix form
\begin{equation}
 {\bf H} {\bf c}_n = \epsilon_n {\bf S} {\bf c}_n \label{eq:gs_assumpt},
\end{equation}
one can derive an approximation to the EC force which we call the incomplete basis set corrected force.
Using the ground state assumption [Eq. (\ref{eq:gs_assumpt})] and the relation between the matrices ${\bf
D}_a$ and ${\bf S}$,
\begin{equation}
 \frac{\partial {\bf S}}{\partial {\bf R}_a} ={\bf D}_a + {\bf D}_a^{\ast},
\end{equation}
one can derive the IBSC force from the EC force [Eq. (\ref{eq:f_econs})]
\begin{equation}
 \mathbf{F}_a^{\text{IBSC}} = -\frac{\partial E_{\text{el}}}{\partial {\bf R}_a} +\sum_n {\bf c}_n^{\ast}
\epsilon_n \frac{\partial
{\bf S}}{\partial {\bf R}_a} {\bf c}_n. \label{eq:f_ibsc}
\end{equation}

Using the above expression, the time derivative of the total Hamiltonian reads as
\begin{equation}
 \frac{d \cal H}{dt}^{\text{IBSC}} = \sum_{n,a} {\bf v}_a \cdot {\bf c}_n^{\ast}(
\epsilon_n \frac{\partial {\bf S}}{\partial {\bf R}_a} - ({\bf H} {\bf S}^{-1} {\bf D}_a + c.c.)) {\bf c}_n.
\label{eq:dhdt_ibs}
\end{equation}
For a linear system, this should be essentially zero because the vectors ${\bf c}_n$ are the eigenstates of
the Hamiltonian matrix ${\bf H}$. For a nonlinear system, however, the time derivative is generally non-zero
because the difference between the gradient of ${\bf S}$ and the ${\bf H} {\bf S}^{-1} {\bf D}_a + c.c.$ terms
might cause fluctuations in the total energy. Especially in the case of nonadiabatic processes, the ground
state assumption [Eq. (\ref{eq:gs_assumpt})] might yield inadequate results in terms of the total energy
conservation. The IBSC force has been suggested for Ehrenfest MD in a finite
basis in literature \cite{Marx2009}.

\subsubsection{The Hellmann-Feynman force}

The most traditional way of treating the atomic forces is the Hellmann-Feynman theorem, i.e., the force
acting on atom $a$ is calculated by taking the expectation value of the gradient of the electronic
Hamiltonian operator with respect to the atom $a$. Moreover, for the sake of consistency (i.e. no
wavefunction gradients), we leave out the gradient of the nonlinear term $\hat{H}'$ in the Hamiltonian
operator. The HF force is an approximation to the IBSC force [Eq. (\ref{eq:f_ibsc})] as it can be derived by
neglecting the gradients of the basis functions with respect to the atomic positions. Therefore,
the HF force in our calculations reads as
\begin{eqnarray}
 {\bf F}_a^{\text{HF}} &= -\sum_n \braket{\psi_n | \frac{\partial
\hat{H}^0}{\partial {\bf R}_a} | \psi_n}. \label{eq:f_hf}
\end{eqnarray}
Now the time derivative of the total Hamiltonian $\mathcal{H}$ [Eq. (\ref{eq:detot_dt})] can be written as
follows
\begin{eqnarray}
 \frac{d \mathcal{H}}{d t}^{\text{HF}} =&& \sum_a {\bf v}_a \cdot \frac{\partial E'}{\partial {\bf R}_a} +
\sum_{n,a}
{\bf v}_a \cdot {\bf c}_n^{\ast}
({\bf \tilde{H}}^{0,a} + {{\bf \tilde{H}}}^{{0,a}^{\ast}}) {\bf c}_n  \nonumber\\
&&- \sum_{n,a} {\bf v}_a \cdot  {\bf c}_n^{\ast} ({\bf H} {\bf S}^{-1} {\bf D}_a + c.c.) {\bf c}_n,
\label{eq:etot_hf}
\end{eqnarray}
where the matrix ${\bf \tilde{H}}^{0,a}$ reads as
\begin{equation}
 \tilde{H}^{0,a}_{ij} = \braket{\chi_i | \hat{H}_0 | \frac{\partial \chi_j}{\partial
{\bf R}_a}}.
\end{equation}
Clearly, Eq. (\ref{eq:etot_hf}) shows that the HF force will not conserve the total energy because the
terms do not cancel out each others in general. However, for position-independent basis sets the HF force
works well, which is the case, e.g., in the Troullier-Martins pseudopotential-based Octopus package
\cite{Andrade2009}. Pseudopotentials on a position-independent basis also simplify
the electronic part of the Ehrenfest MD equations [Eq. (\ref{eq:td_se_gen})] as the ${\bf P}$ term is zero. In
addition to the Octopus package, the HF approach is used in the SIESTA package \cite{Meng2008} which is based
on the LCAO basis set.

\subsection{Time propagation of the electron-ion system} \label{ssec:time_propagation}

In order to carry out actual simulations, a propagation algorithm for the
quantum-classical system [Eqs. (\ref{eq:td_se_gen}) and (\ref{eq:newton_gen})] is required. First, we use
the following splitting for the propagation of coupled electrons and ions
\begin{eqnarray}
 U_{N,e}(t + &&\Delta t, t) = U_N (t + \frac{\Delta t}{2}, t) U_e (t + \Delta
t, t) \nonumber\\
\times&& U_N (t + \Delta t, t + \frac{\Delta t}{2}) + {\cal O}(\Delta
t^3),\label{eq:split}
\end{eqnarray}
where the propagator for the nuclei, $U_N$, is the standard velocity Verlet \cite{Verlet1967},
while the electronic states ($U_e$) are propagated using the so-called
semi-implicit Crank Nicholson (SICN) method \cite{Castro2004, Walter2008}. The idea of the method is to first
approximate the Hamiltonian matrix ${\bf H}$ to be constant during the time step and solve the following
linear equation to obtain the predicted future electronic states ${\bf c}_n^{\text{pred}}$
\begin{eqnarray}
 &&\left[{\bf S}+ i\frac{\Delta t}{2} ({\bf H}(t) + {\bf P})\right]
{\bf c}_n^{\text{pred}}(t+\Delta t) \nonumber\\
&&= \left[{\bf S} - i \frac{\Delta t}{2} ({\bf H}(t) + {\bf P})\right]
{\bf c}_n (t) + {\cal O} (\Delta t^2). \label{eq:cn_pred}
\end{eqnarray}
Then, the predicted future Hamiltonian matrix, based on ${\bf c}_n^{\text{pred}}$, is used for calculating the
Hamiltonian matrix in the middle of the time step, ${\bf H}(t+\Delta t/2) = \frac{1}{2}({\bf H}(t) + {\bf
H}^{\text{pred}}(t + \Delta t)) + \mathcal{O} (\Delta t^2)$. Now, using the notation ${\bf H}_{1/2} = {\bf H}
(t + \Delta t/2)$, the final, propagated states ${\bf c}_n(t + \Delta t)$ can be obtained from
\begin{eqnarray}
&& \left[{\bf S} + i \frac{\Delta t}{2}({\bf H}_{1/2} +
{\bf P})\right]{\bf c}_n(t+ \Delta t) \nonumber\\
&&= \left[{\bf S} - i \frac{\Delta t}{2} ({\bf H}_{1/2} +
{\bf P})\right] {\bf c}_n(t) + \mathcal{O} ( \Delta t^3). \label{eq:cn_corr}
\end{eqnarray}

\section{Ehrenfest MD in a finite element basis} \label{sec:efmd_fem}

\subsection{The model system} \label{ssec:fem_model}

\begin{figure}
\includegraphics{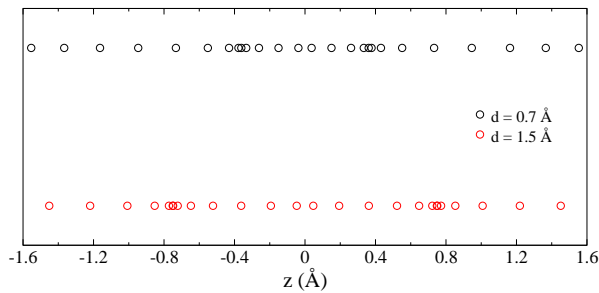}% Here is how to import EPS art
\caption{\label{fig:fem_grids} Non-uniform grid used for the 1D model system. The grid is shown for two
different interatomic distances $d = 0.7$ {\AA} and $d = 1.5$ {\AA}. For illustratory purposes, the small
number of FEM basis functions $N=50$ is used.}
\end{figure}

In order to investigate different forces and their effect on the total energy conservation, we study a
two-atom system using the finite element method (FEM). The basis functions depend explicitly on the atomic
positions. We restrict our calculations to one dimension in order to minimize the errors arising from
discretization. In order to create a
position-dependent basis, we define a mapping $f$ from the uniform grid to a non-uniform grid as follows
\begin{equation}
 f(x) = x \left\vert \frac{x}{L} \right\vert^{\kappa - 1},
\end{equation}
where $L$ is the half-width of the grid and $\kappa > 1$ is a parameter. This mapping creates a grid that is
more dense near the origin and more sparse near the boundaries. Then, we define another mapping $g$
\begin{eqnarray}
 z = g(f(x)) =&& f(x) - [(f(x) - R_1) e^{-\eta (f(x)-R_1)^2} \nonumber\\
\times&& (1-e^{-\nu (f(x) - R_2)^2}) \nonumber\\
+&& (f(x) - R_2) e^{-\eta (f(x)-R_2)^2} \nonumber \\
\times&&(1-e^{-\nu (f(x) - R_1)^2})],
\end{eqnarray}
where $R_1$ and $R_2$ are the atomic positions, and $\eta, \nu > 0$ are parameters. With the above equation,
the points on the uniform grid are mapped onto a non-uniform grid that is more dense around the atoms. 
The piecewise linear FEM basis functions $\chi_k$ are then defined on the non-uniform grid as follows
\begin{equation}
 \chi_k (z) = \left\{ \begin{array}{ll}
 \frac{z - z_{k-1}}{z_k - z_{k-1}} &\mbox{ if $z \in [z_{k-1},z_k]$,} \\
 \frac{z_{k+1} - z}{z_{k+1} - z_{k}}  &\mbox{ if $z \in [z_k, z_{k+1}]$,} \\
  0 &\mbox{ otherwise.}
       \end{array} \right.
\end{equation}
Figure \ref{fig:fem_grids} illustrates the non-uniform grid for two different interatomic distances, $d = 0.7$
{\AA} and $d = 1.5$ {\AA}.
One can clearly see that the grid is more dense close to and between the atoms, and in turn the sparsity
of the grid increases as one approaches its boundaries.

The two-atom system, in our calculations, is defined by the Gross-Pitaevskii (GP)-type (see, for example,
Ref. \onlinecite{Bao2003} and references therein) energy functional
\begin{equation}
 E^{\text{GP}}[\rho] = T_s [\rho] + \int V_{\text{ext}} \rho dz + \frac{\gamma}{2}
\int \rho^2 dz \label{eq:gp_func},
\end{equation}
where $\rho$ is the electronic density, $T_s$ is the kinetic energy of the non-interacting electrons,
$V_{\text{ext}}$ is the
external potential and $\gamma \geq 0$ is a parameter. The external potential in our calculations includes the
nucleus-nucleus (nn) and electron-nucleus (ne) interactions. In order to prevent the potential from diverging
at the nuclei, we use the following, so called soft Coulomb potentials
\begin{eqnarray}
 && V_{\text{ext}}(x) = V_{\text{ne, soft}} + V_{\text{nn, soft}} \nonumber \\
&&= - \frac{a_1}{\sqrt{(x-R_1)^2 + \alpha_1}} -
\frac{a_2}{\sqrt{(x-R_2)^2 + \alpha_1}} \nonumber \\
+&& \frac{\beta}{\sqrt{(R_1-R_2)^2 + \alpha_2}}, 
\end{eqnarray}
where $a_1, a_2, \alpha_1, \alpha_2, \beta > 0$ are parameters. By varying the GP functional
[Eq. (\ref{eq:gp_func})] with respect to the density, we obtain the following Hamiltonian operator
\begin{eqnarray}
 \hat{H} =&& \hat{T} + V_{\text{ne, soft}} + V_{\text{nn, soft}} + \gamma \rho. \label{eq:h_1dfem}
\end{eqnarray}
Now, the matrix ${\bf H}'$ [Eq. (\ref{eq:hprime})] reads as
\begin{equation}
  H'_{ij} = \gamma \braket{\chi_i | \rho |\chi_j} = \gamma \sum_n {\bf c}_n^{\ast} {\bf
A}_{ij} {\bf c}_n \label{eq:hprime_fem},
\end{equation}
where the tensor ${\bf A}$ is defined as
\begin{equation}
 A_{ij}^{kl} = \int \chi_i \chi_j \chi_k \chi_l dz.
\end{equation}
Having defined the model system, the forces can be computed according to the formalism presented in Sec.
\ref{ssec:gen_td}, and the time propagation proceeds as presented in Sec. \ref{ssec:time_propagation}.
%Using equation (\ref{eqn:hprime_fem}), we can compute the matrix elements of the
%ime derivative of $\bm{H}'$
%\begin{eqnarray}
% \frac{\partial H'}{\partial t}_{ij} =&& \sum_n (\frac{\partial
%\mathbf{c}_n^{\ast}}{\partial t} \bm{A}_{ij} \mathbf{c}_n + c.c). \nonumber\\
%=&& i \sum_n \bm{c}_n^{\ast} ([(\bm{H} + \bm{P}^{\ast})\bm{S}^{-1} - c.c.]
%\bm{A}_{ij}) \bm{c}_n
%\end{eqnarray}

\subsection{Results for the model system}

In order to study the different forces presented in Sec. \ref{ssec:gen_td}, we present the implementation
of the FEM system described in Sec. \ref{ssec:fem_model}. In the example calculations, we use
the following, reasonable set of parameters $a_1 = 1, a_2 = 3, \alpha_1 = 0.1, \alpha_2 = 0.01, \eta = \nu =
0.8, \beta = 1.2, \kappa = 1.3 $, and $L = 4.0$ {\AA}. The two lowest
states are occupied. Moreover, the atomic masses are $M_1 = 2000$ a.u. and $M_2 = 5000$ a.u. First, we
study a linear system by setting $\gamma = 0$, and carry out calculations where the system is initially at
the interatomic distance of 1.03 {\AA} (the equilibrium value being 0.69 {\AA}),  after which it evolves
freely for a few femtoseconds. Then, we introduce nonlinearity to the model system by performing
calculations with $\gamma = 0.02$ and $\gamma = 0.2$. The total energy conservation with the HF, IBSC and EC
forces is investigated. Finally, we study the validity of the ground state assumption in the IBSC force by
assigning the kinetic energy of 200 eV to the atoms which are initially at their equilibrium distance.

\subsubsection{Linear system}

\begin{figure}
\includegraphics{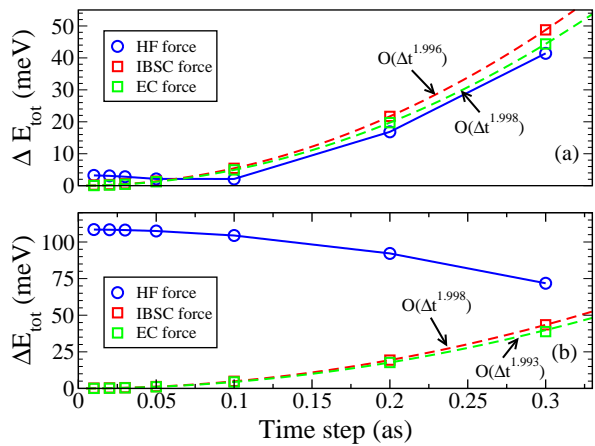}% Here is how to import EPS art
\caption{\label{fig:detot_linf1f2} Decay of the error in the total energy of the linear 1D model system as a
function of the simulation time step.  The results obtained with (a) $N=300$ and (b) $N=50$ FEM basis
functions as well as with the HF, IBSC and EC forces [Eqs. (\ref{eq:f_hf}), (\ref{eq:f_ibsc}) and
(\ref{eq:f_econs}), respectively] are shown. The dashed lines are quadratic fits, while the blue solid line is
just a guide to the eye. The results obtained with ${\bf P} = 0$ do not fit within the scales of the figure. }
\end{figure}

First, it is instructive to demonstrate the effect of the ${\bf P}$ term on the conservation of the norm of
the electronic states and
subsequently total energy conservation. One might assume that neg\-lecting this term would merely cause
small fluctuations in the total energy. In order to investigate the validity of this
assumption, we simulate the two-atom system for $T = 3$ fs using $N = 300$ FEM basis functions and different
time steps. Figure \ref{fig:detot_linf1f2} shows the maximum fluctuation in the total energy within the
simulation time, 
\begin{equation}
 \Delta E_{\text{tot}} = \max\lbrace\vert E_{\text{tot}} (t) - E_{\text{tot}}
(0) | : t \in [0,T]\rbrace,
\end{equation}
as a function of the simulation time step when the ${\bf P}$ term is not neglected. 
Clearly, the difference between the IBSC and EC forces is negligible as both of them
result in a quadratic decay of the error in the total energy. This result confirms the assumption presented in
Sec. \ref{ssec:ibsc} that for a linear system the IBSC force should be good enough. Moreover, in Fig.
\ref{fig:detot_linf1f2}(a), the total energy error with the HF force is roughly as small as that with the EC
and IBSC forces. This is not surprising as the basis is quite large ($N$ = 300). In the case of $N$ = 50,
however, the size of the basis clearly affects the total energy conservation as the error is over 100 meV
(Fig. \ref{fig:detot_linf1f2}(b)). Interestingly, in Fig. \ref{fig:detot_linf1f2}(b), the change in the
total energy for the HF force seems to decrease as the time step is increased. However, this is simply due to
cancellation of errors in this particular case, and for larger time steps than shown in Fig.
\ref{fig:detot_linf1f2}(b) the energy change begins to increase again. Exactly this behavior is seen in Fig.
\ref{fig:detot_linf1f2}(a), but on a different scale.

By setting ${\bf P} = 0$, the behaviour of the total energy changes radically as the norm of the electronic
states is no longer conserved. We find that the maximum fluctuation in the total energy is 28.3 eV which is
580 \% of the maximum kinetic energy of the molecule. Furthermore, the norm of the electronic states
differs from the value of 2 electrons by as much as 0.128 electrons. Thus, the ${\bf P}$ term
has a very significant effect on the total energy conservation and must not be neglected.

\begin{figure}
\includegraphics{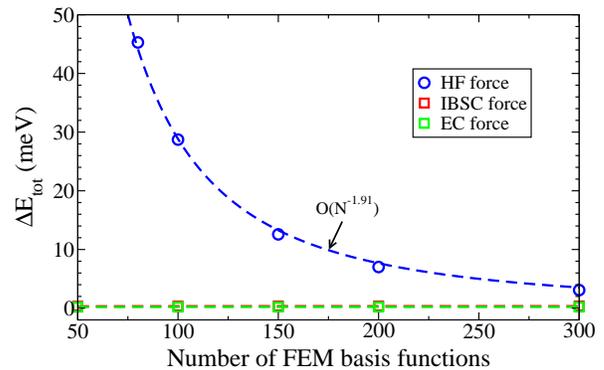}% Here is how to import EPS art
\caption{\label{fig:detot_dn} Error in the total energy of the linear 1D model system with respect to the
number of FEM basis functions. The results obtained with the HF, IBSC and EC forces [Eqs. (\ref{eq:f_hf}),
(\ref{eq:f_ibsc}) and (\ref{eq:f_econs}), respectively] are shown. The time step used for the calculations is
$\Delta t =$ 0.05 as.}
\end{figure}

Finally, we study the validity of the HF theorem by comparing the total energy error obtained with the HF
force to that obtained with the IBSC and EC forces as a function of the size of the basis set. These results
are shown in Fig. \ref{fig:detot_dn}.
We observe that the HF force works succificiently well for $N \geq 100$ basis functions. On the other hand,
Fig. \ref{fig:detot_dn} shows that even in a practically adiabatic case, a small basis might cause significant
fluctuations in the total energy. Thus, one has to be careful with the number of basis functions in the
calculations when using the HF force in a position-dependent basis set. In order to eliminate the
possibility of total energy fluctuations due to the insufficient number of basis functions, one must use
either the IBSC or the EC force.

\subsubsection{Nonlinear system}
Next, having studied the behaviour of the 1D model system in the linear case, we turn our attention to the
nonlinear case by carrying out calculations with a non-zero value of the nonlinearity parameter $\gamma$
[Eq. (\ref{eq:h_1dfem})].

\begin{figure}
\includegraphics{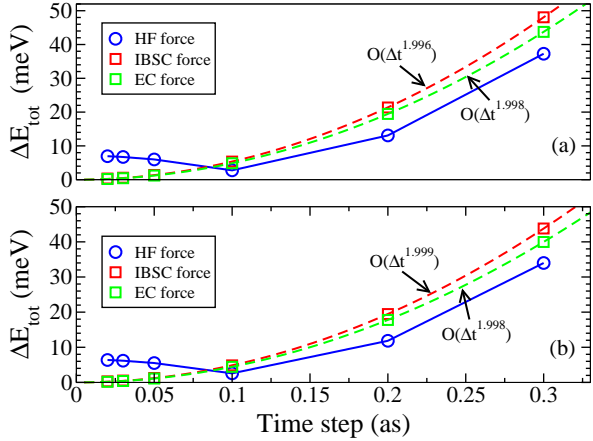}% Here is how to import EPS art
\caption{\label{fig:detot_dt_nl} Error in the total energy of the nonlinear 1D model system as a
function of the simulation time step. The results obtained with the HF, IBSC and EC
forces [Eqs. (\ref{eq:f_hf}), (\ref{eq:f_ibsc}) and (\ref{eq:f_econs}), respectively] and two different
degrees of nonlinearity $\gamma = 0.02$ (a) and $\gamma = 0.2$ (b) are compared. The dashed lines are
quadratic fits, while the blue solid line is just a guide to the eye. }
\end{figure}

First, similarly to the linear case, we perform calculations where the initial atomic separation is $d = 1.03$
{\AA} and investigate the total energy conservation. The total simulation time is $T = 3$ fs, and $N = 200$
FEM basis functions are used in all the calculations. Figure \ref{fig:detot_dt_nl} shows the decay of the
error in the total energy with respect to the simulation time step.
As in the linear case, both the IBSC and the EC forces result in a quadratically decaying total
energy error with respect to the time step. Comparing the results shown in Figs.
\ref{fig:detot_dt_nl}(a) and \ref{fig:detot_dt_nl}(b), the errors with $\gamma = 0.02$ and
$\gamma = 0.2$ appear to be almost identical. Furthermore, the HF force works well with $N = 200$ basis
functions as the total energy error due to the finite basis is only about 7 meV. 

\begin{figure}[!htbp]
\includegraphics{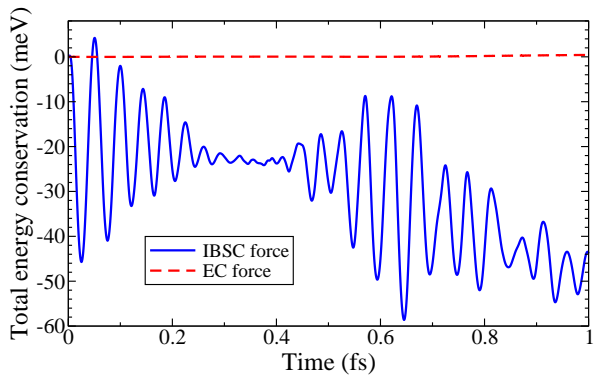}% Here is how to import EPS art
\caption{\label{fig:etot_f1f2ek200} Total energy conservation of the nonlinear 1D model system as a function
of the simulation time. The results obtained with the IBSC and EC forces [Eqs. (\ref{eq:f_ibsc}) and
(\ref{eq:f_econs}), respectively] are compared. The initial kinetic energy of the system is $E_k = 200$ eV. }
\end{figure}

In order to observe a significant difference between the IBSC and EC forces, we have to increase the kinetic
energy of the system. For this reason, we carry out calculations where the atoms are initially in equilibrium.
We then assign the fairly high initial kinetic energy of $E_k = 200$ eV to the atoms, the value of the
nonlinearity parameter being $\gamma = 0.2$. The total energy conservation is investigated within the
simulation time $T = 1$ fs using the time step $\Delta t$ = 0.02 as. Total energy curves for the IBSC
and EC forces are presented in Fig.
\ref{fig:etot_f1f2ek200}.
According to the figure, the EC force conserves the total energy very well within the simulation time of 1
fs. The IBSC force, in contrast, causes clear fluctuations in the total energy -- the error is roughly two
orders of magnitudes larger than that for the EC force. Moreover, instead of fluctuating around a constant
value, the total energy appears to drift downwards as a function of time. Thus, the EC force appears to be a
much better choice for simulations with energetic ions.

\section{Ehrenfest MD within the PAW formalism} \label{sec:efmd_gpaw}

\subsection{Theoretical framework}
Ehrenfest MD within the PAW formalism is similar to the finite basis formalism presented in Sec.
\ref{sec:methodology}. The first notable difference is that within the PAW method, we actually have a basis
defined by two different functions, the projectors $\tilde{p}_i^a$ and the pseudo partial waves
$\tilde{\phi}_i^a$, which fulfill
\begin{equation}
 \braket{\tilde{p}_{i_1} | \tilde{\phi}_{i_2}} = \delta_{i_1, i_2},
\end{equation}
where $i$ is a multi-index consisting of the quantum numbers $l$, $m$ and $n$.
The second notable difference is that the dependency on the atomic positions arises from the
position-dependent PAW transformation operator $\hat{\cal T}$. Similarly to the finite basis set case, there
appears an additional
term due to the moving basis set in the TDDFT-equivalent of the single-particle TD
Schr{\"o}dinger equation [Eq. (\ref{eq:td_se})]. We start from the all-electron TDKS equation,
\begin{equation}
 i \frac{\partial \psi_n}{\partial t} = \hat{H} \psi_n, \label{eq:tdks_ae}
\end{equation}
by applying the PAW transformation,
\begin{equation}
 \psi_n = \hat{\cal T} \tilde{\psi}_n,
\end{equation}
between the all-electron KS wavefunctions $\psi_n$ and the KS pseudo wavefunctions $\tilde{\psi}_n$. Then, the
all-electron TDKS equation [Eq. (\ref{eq:tdks_ae})] is operated from the left by the adjoint of
the PAW transformation operator, $\hat{\mathcal{T}}^{\dagger}$. Subsequently, we arrive at the following
PAW-transformed TDKS equation
\begin{equation}
 i \tilde{S} \frac{\partial \tilde{\psi}_n}{\partial t} = (\tilde{H} +
\tilde{P}) \tilde{\psi}_n, \label{eq:tdks_paw}
\end{equation}
where $\tilde{S}$ is the PAW overlap operator, and $\tilde{H}$ is the PAW Hamiltonian operator. The
$\tilde{P}$ term, which corresponds to the ${\bf P}$ matrix
presented in Sec. \ref{ssec:gen_td}, reads as
\begin{equation}
 \tilde{P} = -i \hat{\cal T}^{\dagger} \frac{\partial \hat{\cal T}}{\partial t}.
\end{equation}
It takes into account the time evolution of the PAW transformation operator
in TDDFT-based quantum-classical MD simulations. Qian et al.\cite{Qian2006} derived the following expression
for this term
\begin{eqnarray}
 \tilde{P} =&& - i \sum_a {\bf v}_a \cdot (1 + \hat{t}_a^{\dagger})\frac{\partial}{\partial {\bf R}_a}(1 +
\hat{t}_a) \nonumber\\
=&& -i \sum_a {\bf v}_a \cdot {\bf \hat{D}}_a, \label{eq:p_term}
\end{eqnarray}
where $\hat{t}_a = \sum_i (\ket{\phi_i^a} - \ket{\tilde{\phi}_i^a})
\bra{\tilde{p}_i^a}$ is a projection operator belonging to atom $a$, and we have defined the operator ${\bf
\hat{D}}_a$ in the spirit of the formalism presented in Sec. \ref{sec:methodology}. Moreover, Eq.
(\ref{eq:p_term}) only holds if the overlap between the PAW augmentation spheres is zero. In practice,
however, Eq. (\ref{eq:p_term}) turns out to work well even in the case of overlapping augmentation
spheres as long the overlap is not significant. The operator ${\bf \hat{D}}_a$ can be written in the following
form [Appendix \ref{ap:p_sym}]
\begin{eqnarray}
 {\bf \hat{D}}_a = &&\sum_{i_1,i_2} [\ket{\tilde{p}_{i_1}^a} O^a_{i_1,i_2}
\bra{\frac{\partial \tilde{p}_{i_2}^a}{\partial {\bf R}_a}} \nonumber \\
&&+ \ket{\tilde{p}_{i_1}^a} ( \braket{\phi_{i_1}^a | \frac{\partial
\phi_{i_2}^a}{\partial {\bf R}_a}} - \braket{\tilde{\phi}_{i_1}^a |
\frac{\partial \tilde{\phi}_{i_2}^a}{\partial {\bf R}_a}})
\bra{\tilde{p}_{i_2}^a}]. \label{eq:d_pawform}
\end{eqnarray}
The matrix elements $O^a_{i_1, i_2}$ describe the overlap between the all-electron and pseudo partial
waves
\begin{equation}
 O^a_{i_1, i_2} = \braket{\phi^a_{i_1} | \phi^a_{i_2}} -
\braket{\tilde{\phi}^a_{i_1} | \tilde{\phi}^a_{i_2}}.
\end{equation}

The next task is to derive a computable expression for the force similar to the EC force [Eq.
(\ref{eq:f_econs})]. Using the same reasoning as in Sec. \ref{sec:methodology}, i.e., the conservation of
the total energy, we replace the IBSC force expression used for ground state calculations in the GPAW
package \cite{Mortensen2005, Enkovaara2010},
\begin{equation}
 {\bf F}^{\text{IBSC}}_a = -\frac{\partial E_{\text{el}}}{\partial {\bf R}_a} + \sum_n f_n
\epsilon_n \braket{\tilde{\psi}_n | \frac{\partial \tilde{S}}{\partial
{\bf R}_a}| \tilde{\psi_n}},\label{eq:gpaw_gsforce}
\end{equation}
where $f_n$ is the occupation number of state $n$, with the more general expression
\begin{equation}
 {\bf F}^{\text{EC}}_a = -\frac{\partial E_{\text{el}}}{\partial {\bf R}_a} + \sum_n f_n
\braket{\tilde{\psi}_n | {\bf \hat{D}}_a^{\dagger} \tilde{S}^{-1} \tilde{H} + c.c. |
\tilde{\psi}_n}. \label{eq:gpaw_tdforce}
\end{equation}
By defining $\tilde{g}_n = \tilde{S}^{-1} \tilde{H} \tilde{\psi}_n$ and the vector-valued matrix elements
\begin{equation}
 \bm{\Delta}_{i_1,i_2}^a = \braket{\phi^a_{i_1} | \frac{\partial
\phi^a_{i_2}}{\partial {\bf R}_a}} - \braket{\tilde{\phi}^a_{i_1} | 
\frac{\partial \tilde{\phi}^a_{i_2}}{\partial {\bf R}_a}},
\end{equation}
we arrive at the following equation
\begin{eqnarray}
  \braket{\tilde{\psi}_n | {\bf \hat{D}}_a^{\dagger} \tilde{S}^{-1} \tilde{H} |
\tilde{\psi}_n} =&& \sum_{i_1,i_2} [ \braket{\tilde{\psi}_n
| \frac{\partial \tilde{p}^a_{i_2}}{\partial {\bf R}_a}} O_{i_1,i_2}^a
\braket{\tilde{p}^a_{i_1} | \tilde{g}_n} \nonumber \\
+&& \braket{\tilde{\psi}_n | \tilde{p}^a_{i_2}} \bm{\Delta}_{i_1,i_2}^a
\braket{\tilde{p}^a_{i_1} | \tilde{g}_n} ]. \label{eq:wfs_corr}
\end{eqnarray}
With the EC force expression for GPAW and the required force corrections [Eqs. (\ref{eq:gpaw_tdforce}) and
(\ref{eq:wfs_corr})], the atomic forces can be straightforwardly calculated. The matrix elements
$O^a_{i_1,i_2}$ and $\bm{\Delta}_{i_1,i_2}^a$ are calculated on radial grids inside the PAW augmentation
spheres, whereas the terms involving the pseudo wavefunctions are computed on uniform Cartesian grids.

The coupled electron-ion system is propagated using the SICN method [Eqs. (\ref{eq:cn_pred}) and
(\ref{eq:cn_corr})] for the electronic states $\tilde{\psi}_n$ and the velocity Verlet algorithm for the
nuclei in a similar fashion as in the 1D calculations in the FEM basis. However, unlike in one dimension, the
calculation of the inverse PAW overlap operator $\tilde{S}^{-1}$ is not trivial. We use two different
approaches for this. In the first approach we assume that the inverse overlap operator can be written as
\begin{equation}
 \tilde{S}^{-1} = 1 + \sum_a \sum_{i_1,i_2} \ket{\tilde{p}^a_{i_1}} C_{i_1,i_2}^a
\bra{\tilde{p}^a_{i_2}} \label{eq:invapr1},
\end{equation}
where $C_{i_1,i_2}^a$ are the inverse overlap coefficients. They can be obtained
from the linear equation resulting from the requirement
\begin{equation}
 \tilde{S}^{-1} \tilde{S} = 1 \label{eq:invapr2},
\end{equation}
in which we also assume that there is no overlap between the PAW augmentation spheres. For this reason, this
approach will likely produce unsatisfactory results if the overlap is non-zero. In the second approach, we
obtain the required $\tilde{S}^{-1} \tilde{H} \tilde{\psi}_n$ terms by solving the linear equations
\begin{equation}
 \tilde{S} \tilde{x} = \tilde{H} \tilde{\psi}_n,
\end{equation}
iteratively using the conjugate gradient (CG) method. This approach is generally more accurate than the
app\-roximative inverse method [Eqs. (\ref{eq:invapr1}) and (\ref{eq:invapr2})] and less sensitive to PAW
overlap and numerical errors due to a large grid spacing. This is illustrated by Table \ref{tbl:h2cnh2}, in
which the approximative inverse method is very sensitive to the grid spacing used in the calculations.
However, being an iterative approach, the CG method will likely consume more computational resources than the
approximative method. 

\subsection{Calculations for small and medium-sized molecules}

In order to test our PAW based Ehrenfest MD method, we study the dynamics of small and medium sized
molecules both in adiabatic and in nonadiabatic cases. The adiabatic cases include the vibration of the NaCl
molecule, and the rotation of the H$_2$C = NH$_2^+$ molecule
about its internal axis with a small initial kinetic energy. Nonadiabatic effects, then, are studied in the
case of H$_2$C = NH$_2^+$ with a high initial kinetic energy, and also in the hydrogen bombardment of the
C$_{40}$H$_{16}$ molecule. We use the LDA exchange-correlation functional \cite{Perdew1992} in all the
calculations. The grid spacing is $h = 0.2$ {\AA} unless specified otherwise.

\subsubsection{Vibration of the NaCl molecule}

\begin{figure}[!htb]
\includegraphics{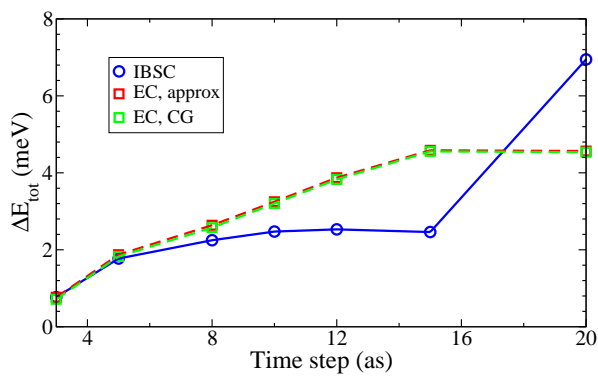}% Here is how to import EPS art
\caption{\label{fig:nacl_etot} Decay of the error in the total energy of the NaCl molecule as a function
of the simulation time step. The results are shown for the IBSC [Eq. (\ref{eq:gpaw_gsforce})] and the EC [Eq.
(\ref{eq:gpaw_tdforce})] forces. Approx denotes the approximative method for calculating the inverse overlap
operator [Eqs. (\ref{eq:invapr1}) and (\ref{eq:invapr2})], and CG corresponds to the conjugate gradient
method. The lines are just a guide to the eye. }
\end{figure}

First, we study the total energy conservation of a simple dimer, the NaCl molecule. The simulations begin
from equilibrium ($d_{\text{eq}}$ = 2.36 {\AA}) with the initial kinetic energy of 2 eV. The period
of the NaCl vibration calculated with the BOMD is $T_{\text{BOMD}} = 181.6$ fs. With this low initial kinetic
energy, the vibration is almost adiabatic, and subsequently the period obtained with the Ehrenfest MD
($\Delta t$ = 8 as, IBSC force) is very close to the BOMD result, $T_{\text{EF}} = 181.4$ fs. This is not
surprising as the Ehrenfest MD should reduce to the BOMD in the case of adiabatic processes. We investigate
the difference between the IBSC and EC forces in terms of the total energy conservation, using both the
approximative and the CG inverse method for calculating the inverse overlap operator. The decay of the error
in the total energy as a function of the simulation time step is shown in Fig. \ref{fig:nacl_etot}.
For time steps below 10 as, there is little difference between the IBSC and EC forces. Actually the total
energy error with the IBSC force is slightly smaller than that with the EC force, thus
rendering the EC force unnecessary in this adiabatic case. Moreover, the two methods used for calculating the
inverse overlap operator give pratically identical results. 

\subsubsection{Rotation of the H$_2$C = NH$_2^+$ molecule}

Next, we turn our attention to simulating nonadiabatic dynamics. The dynamics of the H$_2$C = NH$_2^+$
molecule
has been studied with both the Hartree-Fock based Ehrenfest dynamics \cite{Li2005} and the
trajectory surface hopping method \cite{Tapavicza2007}. We study the rotation of this molecule about its
internal axis by carrying out Ehrenfest MD calculations with two different initial torsional kinetic energies,
$E_k =$ 1.5 eV and $E_k$ = 10 eV. In order to investigate the nonadiabaticity in our simulations, we also
carry out PAW-based BOMD calculations for both initial kinetic energies. Based on the results presented in
Ref. \onlinecite{Li2005}, the rotation is expected to be adiabatic with $E_k$ = 1.5 eV, whereas with $E_k$ =
10 eV we expect the Ehrenfest MD results to clearly differ from the BOMD results. The molecule is initially in
the planar equilibrium geometry. For both initial kinetic energies, we carry out calculations using
two different time steps, $\Delta t = $ 2 and 5 as. Similarly to the calculations for the NaCl molecule, the
IBSC and the EC forces are applied. In the case of the EC force, both the approximative and the CG method are
used for computing the inverse overlap operator . 

\begin{figure}[!htb]
\includegraphics{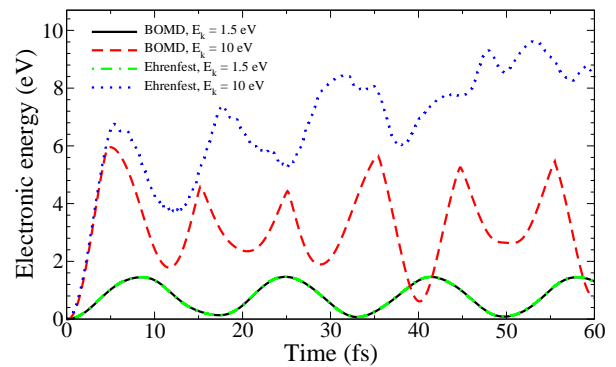}% Here is how to import EPS art
\caption{\label{fig:ch4n_Eel} Time evolution of the PES of the H$_2$C = NH$_2^+$ molecule. The results
obtained with the Ehrenfest MD and the BOMD for low and high initial kinetic energies are compared. In the
BOMD simulations, the time step of 0.1 fs is used. The Ehrenfest MD simulations are performed using the EC
force [Eq. (\ref{eq:gpaw_tdforce})] in conjunction with the CG method for calculating the inverse overlap
operator.}
\end{figure}

The potential energy surfaces obtained from the Ehrenfest MD and the BOMD simulations
for both initial kinetic energies are presented in Fig. \ref{fig:ch4n_Eel}. For the Ehrenfest PES, the EC
force in conjunction with the CG method for calculating $\tilde{S}^{-1}$ is used.
The figure illustrates that the dynamics is nearly adiabatic with $E_k = 1.5$ eV as the Ehrenfest and the
Born-Oppenheimer PES are almost identical. In contrast, with $E_k = 10$ eV, the Ehrenfest PES starts to
deviate rapidly from the BO one, which indicates a significant amount of nonadiabaticity in the dynamics. 

\begin{figure}[!htb]
\includegraphics{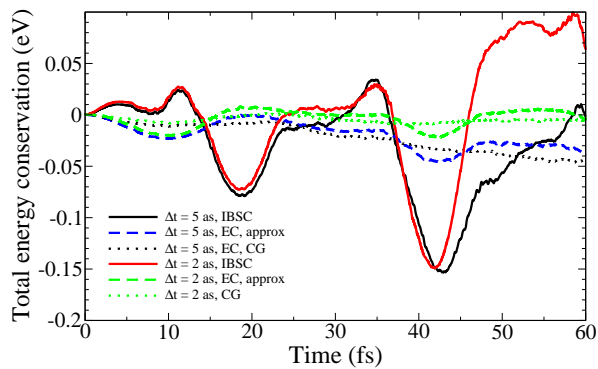}% Here is how to import EPS art
\caption{\label{fig:ch4n_Etot} Conservation of the total energy of the H$_2$C = NH$_2^+$ molecule as a
function of the simulation time. The results obtained with the IBSC and the EC forces [Eqs.
(\ref{eq:gpaw_gsforce}) and (\ref{eq:gpaw_tdforce}), respectively] using two different time steps, $\Delta t
=$ 2 and 5 as, are compared. The initial kinetic energy of the molecule is $E_k$ = 10 eV. Approx and CG
correspond to the approximative [Eqs. (\ref{eq:invapr1}) and
(\ref{eq:invapr2})] and the CG method for calculating the inverse overlap operator, respectively. }
\end{figure}

In the adiabatic case, the total energy is conserved to a few meV even with the IBSC force. With
$E_k = 10$ eV, in contrast, this is no longer the case. The total energy curves
obtained with the different forces and the time steps of 2 and 5 as are shown in Fig. \ref{fig:ch4n_Etot}.
First, we observe that the maximum total energy fluctuation with the IBSC force is roughly the same for both
time steps used in the calculations. Secondly, the EC force in conjunction with the conjugate gradient method
for calculating the inverse overlap operator, yields the best
total energy conservation for both time steps, $\Delta E_{\text{tot}} =$ 45 meV and 9.3 meV for $\Delta t = $
5 and 2 as, respectively. The latter number is very good considering the amount of nonadiabaticity involved in
the dynamics, and it could be further improved by decreasing the time step. Furthermore, with $\Delta t =$ 2
as, the EC force in conjunction with the approximative inverse already improves the total energy conservation
quite significantly compared to the IBSC force. Nevertheless, the error in the total e\-ner\-gy is at least
twice
as high as with the CG inverse. 

Next, we study the effect of the grid spacing on the total energy conservation by carrying out simulations
with $h = 0.15, 0.2, 0.25$ {\AA}. Because the IBSC force works well with the low initial kinetic energy of
$E_k = 1.5$ eV, we only study the more energetic case. The error in the total energy as a function of the grid
spacing is presented in Table \ref{tab:ch4n_Etot_h}.
\begin{table}[!htb]
\caption{\label{tab:ch4n_Etot_h} Maximum total energy fluctuation of the H$_2$C = NH$_2^+$ molecule as
function of the grid spacing $h$. The initial kinetic energy is $E_k$ = 10 eV. The results obtained with the
IBSC and EC forces [Eqs. (\ref{eq:gpaw_gsforce}) and (\ref{eq:gpaw_tdforce}), respectively] are compared. All
the energies are in meV.}
\begin{ruledtabular}
\begin{tabular}{cccc} \label{tbl:h2cnh2}
$h$ (\AA) & $\Delta E_{\text{tot}}^{\text{IBSC}}$ & $\Delta E_{\text{tot}}^{\text{EC,approx}}$ &
$\Delta E_{\text{tot}}^{\text{EC, CG}}$\\
0.15 & 148.5 & 9.95 &  10.01\\
0.2 & 149.19& 23.08& 9.58 \\
0.25 & 151.06 & 228.78  & 8.86 \\
\end{tabular}
\end{ruledtabular}
\end{table}
The accuracy of the approximative inverse increases rapidly as a function of decreasing grid spacing. With $h
= $ 0.25 {\AA}, the IBSC force actually conserves the total energy better than the EC force if the
approximative inverse is used. However, despite the rapid convergence, it is undesireable that
the total energy error has such a strong dependence on the grid spacing. In contrast, the grid spacing has
very
little effect on the results obtained with the CG inverse. For this reason, even though the calculations might
require, depending on the system, 10-20 \% more computational time, the CG inverse should be used for
simulations involving nonadiabatic effects ins\-tead of the approximative inverse. 

\subsubsection{Hydrogen bombardment of the C$_{40}$H$_{16}$ molecule}

\begin{figure}[!htb]
\includegraphics{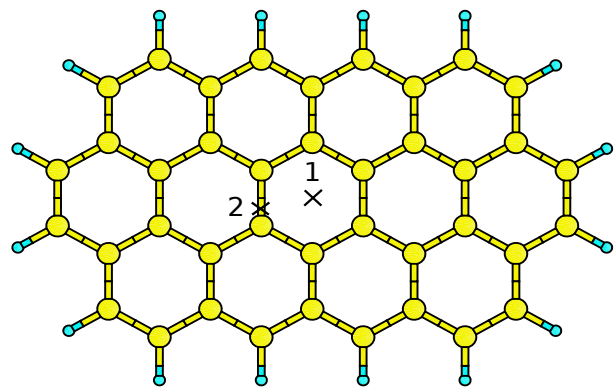}% Here is how to import EPS art
\caption{\label{fig:c40h16_coord} C$_{40}$H$_{16}$ molecule and the two representative trajectories
used in the simulations. }
\end{figure}

As the final test for our PAW-based Ehrenfest MD method, we study the collision of hydrogen with the
graphene-like nanoflake C$_{40}$H$_{16}$. With high enough impact energies, the hydrogen projectile will
invoke electronic excitations in the target, rendering the BOMD approach unusable.
In Ref. \onlinecite{Krasheninnikov2007}, a similar hydrogen ion stopping process was studied in the case of
graphene, for two representative trajectories, 1) center of hexagon and 2) the impact parameter of 0.25 {\AA}
along the C-C bond. The two trajectories are illustrated in Fig. 
\ref{fig:c40h16_coord}.
The EC force in conjunction with the CG method for computing the inverse overlap
operator is used in all the calculations. The time step varies between 1 and 5 as such that $\Delta t$ = 1 as
is used in the simulation with the highest initial projectile
energy, $E_k = 10$ keV and $\Delta t$ = 5 as in that with the lowest initial energy, $E_k$ = 100 eV. 

\begin{figure}[!htb]
\includegraphics{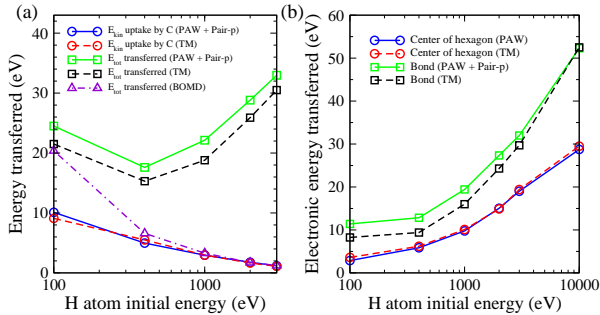}
\caption{\label{fig:c40h16_etransfer} Deposition of the energy into the individual degrees of freedom in the
collision of the H atom with the C$_{40}$H$_{16}$ molecule. (a) Energy transferred to the target C atom, and
the total transferred energy as a function of the H impact energy. The results are shown for the
bond trajectory. Pair-p denotes the pair-potential corrections. (b) Energy transferred into the electronic
degrees of freedom as a function of H impact energy for both trajectories. The lines are just a guide to the
eye. }
\end{figure}

We study the accomodation of the energy transferred into the individual degrees of freedom. Because of
the overlap between the augmentation spheres, the
PAW method underestimates the atomic forces when the interatomic distance is small. Consequently, we use pair-
potential corrections derived from results obtained with FHI-aims \cite{Blum2009} when the distance between
the projectile and the nearest carbon atom is smaller than 0.5
{\AA}. Figure \ref{fig:c40h16_etransfer}(a) shows the total transferred energy and the C recoil energy for the
bond trajectory. Electronic excitations significantly influence the results beyond the impact energy of 400 eV
as the total transferred energy starts to increase. This observation is in agreement with the
Troullier-Martins pseudopotential calculations for graphene in Ref. \onlinecite{Krasheninnikov2007}.
However, despite the qualitative agreement, the quantitative results differ slightly,
which can be attributed to the heavy overlap between the augmentation spheres of the hydrogen projectile and
the target carbon atom. Nevertheless, the agreement is quite good considering that the Eqs. (\ref{eq:p_term})
and (\ref{eq:d_pawform}) used for calculating the
$\tilde{P}$ operator are in principle only correct when the
augmentation spheres do not overlap.

In Fig. \ref{fig:c40h16_etransfer}(b), the energy transferred into the electronic degrees of freedom is
presented. The PAW and TM results are in good agreement for the center of hexagon trajectory,
i.e., when there is no overlap between the PAW augmentation spheres. Thus, our method describes electronic
excitations in a similar fashion as the TM based method. Consequently, it seems that our method works
correctly also in nonadiabatic cases as long as the overlap between the PAW augmentation spheres is not
significant. Even in the case of overlapping augmentation spheres, our method can predict qualitative trends
with the help of pair-potential corrections.

Finally, we summarize the results regarding the total energy conservation in the calculations. First, in all
the simulations for the center of hexagon trajectory, the total energy is conserved to better than 2.7 meV,
which is an excellent result considering that the amount of energy deposited into the electronic degrees of
freedom is of the order of tens of eV. In the case of the bond trajectory, unfortunately, such good numbers
cannot be obtained -- the total energy is conserved to better than 230 meV. This is probably due to the
breakdown of the zero overlap approximation in deriving the $\tilde{P}$ term as it is essential that this
term is correct in order to conserve the total energy. Nevertheless, significant PAW overlap is quite unusual
in Ehrenfest MD applications. Thus, in most cases our method can be expected to conserve the total energy
very well.                                      

\section{Conclusions} \label{sec:discussion}

We have described the implementation of the Ehrenfest molecular dynamics within the time-dependent density
functional theory and the projector augmented-wave method. Moreover, we have studied different Ehrenfest MD
forces using a position-dependent finite element basis in one dimension as well as in 3D with the GPAW
package. In the 1D calculations, the forces were compared by carrying out Ehrenfest MD simulations for a
two-atom system. In the 3D calculations, the dynamics of small and medium sized molecules was studied
both in adiabatic and in nonadiabatic cases. The incomplete basis set corrected and Hellmann-Feynman forces
were found to work succifiently well in adiabatic or nearly adiabatic cases, whereas in clearly nonadiabatic
cases unphysical fluctuations in the
total energy were observed. The total energy conserving force was found to work well also in the nonadiabatic
regime. Finally, the PAW-based Ehrenfest MD results were compared to Troullier-Martins pseudopotential
calculations. From these results, we conclude that our method seems applicable to simulating the nonadiabatic
dynamics of medium sized molecules and beyond as long as the PAW augmentation spheres do not overlap
significantly.

\begin{acknowledgements}
This work has been supported by the Academy of Finland (the Center of Excellence program). The
computational time was provided by the Finnish IT Center for Science (CSC) and the Triton cluster of the
Aalto School of Science. One of the authors (L. L.) acknowledges the support from the French ANR
(ANR-08-CEXC8-008-01). The electronic structure program GPAW is developed in collaboration with
CAMd/Technical University of Denmark, CSC, Department of Physics/University of Jyv{\"a}skyl{\"a}, Institute of
Physics/Tampere University of Technology, and Department of Applied Physics/Aalto University. 
\end{acknowledgements}

\appendix

\section{Symmetric form of the operator ${\bf \hat{D}}_a$} \label{ap:p_sym}

In order to carry out Ehrenfest MD simulations in practice, it is useful to write the $\tilde{P}$ term [Eq.
(\ref{eq:p_term})], in a symmetric form reminiscent of the other observables within the PAW method. This can
be achieved by deriving a symmetric form for its constituent operator ${\bf \hat{D}}_a$ [Eq.
(\ref{eq:d_pawform})]. We start by expanding this operator in terms of the PAW projectors and partial waves
\begin{eqnarray}
&&{\bf \hat{D}}_a = (1+\hat{t}_a^{\dagger})  \frac{\partial \hat{t}_a}{\partial {\bf R}_a} =
\frac{\partial}{\partial
{\bf R}_a} \sum_{i_2} (\ket{\phi_{i_2}^a} - \ket{\tilde{\phi}_{i_2}^a}) \bra{\tilde{p}_{i_2}^a} \nonumber \\
&& + \sum_{i_1} \ket{\tilde{p}_{i_1}^a} (\bra{\phi_{i_1}^a} - \bra{\tilde{\phi}_{i_1}^a}) \nonumber \\
&&\times \frac{\partial}{\partial {\bf R}_a} \sum_{i_2} (\ket{\phi_{i_2}^a} - \ket{\tilde{\phi}_{i_2}^a})
\bra{\tilde{p}_{i_2}^a}.
\end{eqnarray}
Rearranging the terms and adding and substracting a suitable term gives
\begin{eqnarray}
&&{\bf \hat{D}}_a = (1 - \sum_{i_1} \ket{\tilde{p}_{i_1}^a}\bra{\tilde{\phi}_{i_1}^a})
\frac{\partial}{\partial
{\bf R}_a} \sum_{i_2} (\ket{\phi_{i_2}^a} - \ket{\tilde{\phi}_{i_2}^a})
\bra{\tilde{p}_{i_2}^a} \nonumber\\
&&+ \sum_{i_1} \ket{\tilde{p}_{i_1}^a}\bra{\phi_{i_1}^a} \frac{\partial}{\partial {\bf R}_a} \sum_{i_2}
(\ket{\phi_{i_2}^a} - \ket{\tilde{\phi}_{i_2}^a})
\bra{\tilde{p}_{i_2}^a} \nonumber \\
&&+ \sum_{i_1} \ket{\tilde{p}_{i_1}^a}\bra{\tilde{\phi}_{i_1}^a} \frac{\partial}{\partial {\bf R}_a}
\sum_{i_2} (\ket{\phi_{i_2}^a} - \ket{\tilde{\phi}_{i_2}^a}) \bra{\tilde{p}_{i_2}^a} \nonumber\\
&&- \sum_{i_1} \ket{\tilde{p}_{i_1}^a}\bra{\tilde{\phi}_{i_1}^a} \frac{\partial}{\partial {\bf R}_a}
\sum_{i_2} (\ket{\phi_{i_2}^a} - \ket{\tilde{\phi}_{i_2}^a}) \bra{\tilde{p}_{i_2}^a}.
\end{eqnarray}
Using the orthonormality of the projectors and pseudo partial waves, we obtain the following expression
\begin{eqnarray}
 &&{\bf \hat{D}}_a = -\sum_{i_1} \ket{\tilde{p}_{i_1}^a} \bra{\phi_{i_1}^a} \frac{\partial}{\partial
{\bf R}_a} \sum_{i_2} \ket{\tilde{\phi}_{i_2}^a}\bra{\tilde{p}_{i_2}^a} \nonumber\\
&&+ \sum_{i_1} \ket{\tilde{p}_{i_1}^a} \bra{\phi_{i_1}^a}  \frac{\partial}{\partial
{\bf R}_a} \sum_{i_2} \ket{\phi_{i_2}^a}\bra{\tilde{p}_{i_2}^a} \nonumber \\
&&- \sum_{i_1} \ket{\tilde{p}_{i_1}^a} \bra{\tilde{\phi}_{i_1}^a}  \frac{\partial}{\partial
{\bf R}_a} \sum_{i_2} \ket{\phi_{i_2}^a}\bra{\tilde{p}_{i_2}^a} \nonumber\\
&&+ \sum_{i_1} \ket{\tilde{p}_{i_1}^a} \bra{\tilde{\phi}_{i_1}^a} \frac{\partial}{\partial
{\bf R}_a} \sum_{i_2} (\ket{\phi_{i_2}^a} - \ket{\tilde{\phi}_{i_2}^a})\bra{\tilde{p}_{i_2}^a}.
\label{eq:d_sym}
\end{eqnarray}
The first term in Eq. (\ref{eq:d_sym}) is zero due to orthonormality. Combining the remaining terms then
yields
the desired symmetric form of the ${\bf \hat{D}}_a$ operator,
\begin{eqnarray}
 {\bf \hat{D}}_a =&& \sum_{i_1,i_2} (\ket{\tilde{p}_{i_1}^a} O^a_{i_1,i_2}
\bra{\frac{\partial \tilde{p}_{i_2}^a}{\partial {\bf R}_a}} \nonumber \\
&&+ \ket{\tilde{p}_{i_1}^a} ( \braket{\phi_{i_1}^a | \frac{\partial
\phi_{i_2}^a}{\partial {\bf R}_a}} - \braket{\tilde{\phi}_{i_1}^a |
\frac{\partial \tilde{\phi}_{i_2}^a}{\partial {\bf R}_a}})
\bra{\tilde{p}_{i_2}^a}).
\end{eqnarray}

%

%\bibliography{efmd}

\begin{thebibliography}{10}%
\makeatletter
\providecommand \@ifxundefined [1]{%
 \ifx #1\undefined \expandafter \@firstoftwo
 \else \expandafter \@secondoftwo
\fi
}%
\providecommand \@ifnum [1]{%
 \ifnum #1\expandafter \@firstoftwo
 \else \expandafter \@secondoftwo
\fi
}%
\providecommand \enquote [1]{``#1''}%
\providecommand \bibnamefont  [1]{#1}%
\providecommand \bibfnamefont [1]{#1}%
\providecommand \citenamefont [1]{#1}%
\providecommand\href[0]{\@sanitize\@href}%
\providecommand\@href[1]{\endgroup\@@startlink{#1}\endgroup\@@href}%
\providecommand\@@href[1]{#1\@@endlink}%
\providecommand \@sanitize [0]{\begingroup\catcode`\&12\catcode`\#12\relax}%
\@ifxundefined \pdfoutput {\@firstoftwo}{%
 \@ifnum{\z@=\pdfoutput}{\@firstoftwo}{\@secondoftwo}%
}{%
 \providecommand\@@startlink[1]{\leavevmode}%
 \providecommand\@@endlink[0]{}%
}{%
 \providecommand\@@startlink[1]{%
  \leavevmode
  \pdfstartlink
   attr{/Border[0 0 1 ]/H/I/C[0 1 1]}%
   user{/Subtype/Link/A<</Type/Action/S/URI/URI(#1)>>}%
  \relax
 }%
 \providecommand\@@endlink[0]{\pdfendlink}%
}%
\providecommand \url  [0]{\begingroup\@sanitize \@url }%
\providecommand \@url [1]{\endgroup\@href {#1}{\urlprefix}}%
\providecommand \urlprefix [0]{URL }%
\providecommand \Eprint[0]{\href }%
\@ifxundefined \urlstyle {%
  \providecommand \doi [1]{doi:\discretionary{}{}{}#1}%
}{%
  \providecommand \doi [0]{doi:\discretionary{}{}{}\begingroup
  \urlstyle{rm}\Url }%
}%
\providecommand \doibase [0]{http://dx.doi.org/}%
\providecommand \Doi[1]{\href{\doibase#1}}%
\providecommand \selectlanguage [0]{\@gobble}%
\providecommand \bibinfo [0]{\@secondoftwo}%
\providecommand \bibfield [0]{\@secondoftwo}%
\providecommand \translation [1]{[#1]}%
\providecommand \BibitemOpen[0]{}%
\providecommand \bibitemStop [0]{}%
\providecommand \bibitemNoStop [0]{.\EOS\space}%
\providecommand \EOS [0]{\spacefactor3000\relax}%
\providecommand \BibitemShut [1]{\csname bibitem#1\endcsname}%
%</preamble>
\bibitem{Car1985}%
  \BibitemOpen
  \bibfield{author}{%
  \bibinfo {author} {\bibfnamefont{R.}~\bibnamefont{Car}}\ and\ \bibinfo
  {author} {\bibfnamefont{M.}~\bibnamefont{Parrinello}},\ }%
  \bibfield{journal}{%
  \bibinfo {journal} {Phys. Rev. Lett.}\ }%
  \textbf{\bibinfo {volume} {55}},\ \bibinfo {pages} {2471} (\bibinfo {year}
  {1985})\BibitemShut{NoStop}%
\bibitem{Marx2009}%
  \BibitemOpen
  \bibfield{author}{%
  \bibinfo {author} {\bibfnamefont{D.}~\bibnamefont{Marx}}\ and\ \bibinfo
  {author} {\bibfnamefont{J.}~\bibnamefont{Hutter}},\ }%
  \emph{\bibinfo {title} {Ab Initio Molecular Dynamics: Basic Theory and
  Advanced Methods}}\ (\bibinfo {publisher} {Cambridge University Press},\
  \bibinfo {year} {2009})\BibitemShut{NoStop}%
\bibitem{Yabana1996}%
  \BibitemOpen
  \bibfield{author}{%
  \bibinfo {author} {\bibfnamefont{K.}~\bibnamefont{Yabana}}\ and\ \bibinfo
  {author} {\bibfnamefont{G.~F.}\ \bibnamefont{Bertsch}},\ }%
  \bibfield{journal}{%
  \bibinfo {journal} {Phys. Rev. B}\ }%
  \textbf{\bibinfo {volume} {54}},\ \bibinfo {pages} {4484} (\bibinfo {year}
  {1996})\BibitemShut{NoStop}%
\bibitem{Runge1984}%
  \BibitemOpen
  \bibfield{author}{%
  \bibinfo {author} {\bibfnamefont{E.}~\bibnamefont{Runge}}\ and\ \bibinfo
  {author} {\bibfnamefont{E.~K.~U.}\ \bibnamefont{Gross}},\ }%
  \bibfield{journal}{%
  \bibinfo {journal} {Phys. Rev. Lett.}\ }%
  \textbf{\bibinfo {volume} {52}},\ \bibinfo {pages} {997} (\bibinfo {year}
  {1984})\BibitemShut{NoStop}%
\bibitem{Tapavicza2007}%
  \BibitemOpen
  \bibfield{author}{%
  \bibinfo {author} {\bibfnamefont{E.}~\bibnamefont{Tapavicza}}, \bibinfo
  {author} {\bibfnamefont{I.}~\bibnamefont{Tavernelli}},\ and\ \bibinfo
  {author} {\bibfnamefont{U.}~\bibnamefont{Rothlisberger}},\ }%
  \bibfield{journal}{%
  \bibinfo {journal} {Phys. Rev. Lett.}\ }%
  \textbf{\bibinfo {volume} {98}},\ \bibinfo {pages} {023001} (\bibinfo {year}
  {2007})\BibitemShut{NoStop}%
\bibitem{Tully1998}%
  \BibitemOpen
  \bibfield{author}{%
  \bibinfo {author} {\bibfnamefont{J.~C.}\ \bibnamefont{Tully}},\ }%
  \bibfield{journal}{%
  \bibinfo {journal} {Faraday Discuss.}\ }%
  \textbf{\bibinfo {volume} {110}},\ \bibinfo {pages} {407} (\bibinfo {year}
  {1998})\BibitemShut{NoStop}%
\bibitem{Isborn2007}%
  \BibitemOpen
  \bibfield{author}{%
  \bibinfo {author} {\bibfnamefont{C.~M.}\ \bibnamefont{Isborn}}, \bibinfo
  {author} {\bibfnamefont{X.}~\bibnamefont{Li}},\ and\ \bibinfo {author}
  {\bibfnamefont{J.~C.}\ \bibnamefont{Tully}},\ }%
  \bibfield{journal}{%
  \bibinfo {journal} {J. Chem. Phys.}\ }%
  \textbf{\bibinfo {volume} {126}},\ \bibinfo {pages} {134307} (\bibinfo {year}
  {2007})\BibitemShut{NoStop}%
\bibitem{Miyamoto2006}%
  \BibitemOpen
  \bibfield{author}{%
  \bibinfo {author} {\bibfnamefont{Y.}~\bibnamefont{Miyamoto}}, \bibinfo
  {author} {\bibfnamefont{A.}~\bibnamefont{Rubio}},\ and\ \bibinfo {author}
  {\bibfnamefont{D.}~\bibnamefont{Tom{\'{a}}nek}},\ }%
  \bibfield{journal}{%
  \bibinfo {journal} {Phys. Rev. Lett.}\ }%
  \textbf{\bibinfo {volume} {97}},\ \bibinfo {pages} {126104} (\bibinfo {year}
  {2006})\BibitemShut{NoStop}%
\bibitem{Krasheninnikov2007}%
  \BibitemOpen
  \bibfield{author}{%
  \bibinfo {author} {\bibfnamefont{A.~V.}\ \bibnamefont{Krasheninnikov}},
  \bibinfo {author} {\bibfnamefont{Y.}~\bibnamefont{Miyamoto}},\ and\ \bibinfo
  {author} {\bibfnamefont{D.}~\bibnamefont{Tom{\'{a}}nek}},\ }%
  \bibfield{journal}{%
  \bibinfo {journal} {Phys. Rev. Lett.}\ }%
  \textbf{\bibinfo {volume} {99}},\ \bibinfo {pages} {016104} (\bibinfo {year}
  {2007})\BibitemShut{NoStop}%
\bibitem{Meng2008}%
  \BibitemOpen
  \bibfield{author}{%
  \bibinfo {author} {\bibfnamefont{S.}~\bibnamefont{Meng}}\ and\ \bibinfo
  {author} {\bibfnamefont{E.}~\bibnamefont{Kaxiras}},\ }%
  \bibfield{journal}{%
  \bibinfo {journal} {J. Chem. Phys.}\ }%
  \textbf{\bibinfo {volume} {129}},\ \bibinfo {pages} {054110} (\bibinfo {year}
  {2008})\BibitemShut{NoStop}%
\bibitem{Sugino1999}%
  \BibitemOpen
  \bibfield{author}{%
  \bibinfo {author} {\bibfnamefont{O.}~\bibnamefont{Sugino}}\ and\ \bibinfo
  {author} {\bibfnamefont{Y.}~\bibnamefont{Miyamoto}},\ }%
  \bibfield{journal}{%
  \bibinfo {journal} {Phys. Rev. B}\ }%
  \textbf{\bibinfo {volume} {59}},\ \bibinfo {pages} {2579} (\bibinfo {year}
  {1999})\BibitemShut{NoStop}%
\bibitem{Andrade2009}%
  \BibitemOpen
  \bibfield{author}{%
  \bibinfo {author} {\bibfnamefont{X.}~\bibnamefont{Andrade}}, \bibinfo
  {author} {\bibfnamefont{A.}~\bibnamefont{Castro}}, \bibinfo {author}
  {\bibfnamefont{D.}~\bibnamefont{Zueco}}, \bibinfo {author}
  {\bibfnamefont{J.~L.}\ \bibnamefont{Alonso}}, \bibinfo {author}
  {\bibfnamefont{P.}~\bibnamefont{Echenique}}, \bibinfo {author}
  {\bibfnamefont{F.}~\bibnamefont{Falceto}},\ and\ \bibinfo {author}
  {\bibfnamefont{A.}~\bibnamefont{Rubio}},\ }%
  \bibfield{journal}{%
  \bibinfo {journal} {J. Chem. Th. Comp.}\ }%
  \textbf{\bibinfo {volume} {5}},\ \bibinfo {pages} {728} (\bibinfo {year}
  {2009})\BibitemShut{NoStop}%
\bibitem{Blochl1994}%
  \BibitemOpen
  \bibfield{author}{%
  \bibinfo {author} {\bibfnamefont{P.~E.}\ \bibnamefont{Bl{\"{o}}chl}},\ }%
  \bibfield{journal}{%
  \bibinfo {journal} {Phys. Rev. B}\ }%
  \textbf{\bibinfo {volume} {50}},\ \bibinfo {pages} {17953} (\bibinfo {year}
  {1994})\BibitemShut{NoStop}%
\bibitem{Walter2008}%
  \BibitemOpen
  \bibfield{author}{%
  \bibinfo {author} {\bibfnamefont{M.}~\bibnamefont{Walter}}, \bibinfo {author}
  {\bibfnamefont{H.}~\bibnamefont{H{\"{a}}kkinen}}, \bibinfo {author}
  {\bibfnamefont{L.}~\bibnamefont{Lehtovaara}}, \bibinfo {author}
  {\bibfnamefont{M.}~\bibnamefont{Puska}}, \bibinfo {author}
  {\bibfnamefont{J.}~\bibnamefont{Enkovaara}}, \bibinfo {author}
  {\bibfnamefont{C.}~\bibnamefont{Rostgaard}},\ and\ \bibinfo {author}
  {\bibfnamefont{J.~J.}\ \bibnamefont{Mortensen}},\ }%
  \bibfield{journal}{%
  \bibinfo {journal} {J. Chem. Phys.}\ }%
  \textbf{\bibinfo {volume} {128}},\ \bibinfo {pages} {244101} (\bibinfo {year}
  {2008})\BibitemShut{NoStop}%
\bibitem{Walker2007}%
  \BibitemOpen
  \bibfield{author}{%
  \bibinfo {author} {\bibfnamefont{B.}~\bibnamefont{Walker}}\ and\ \bibinfo
  {author} {\bibfnamefont{R.}~\bibnamefont{Gebauer}},\ }%
  \bibfield{journal}{%
  \bibinfo {journal} {J. Chem. Phys.}\ }%
  \textbf{\bibinfo {volume} {127}},\ \bibinfo {pages} {164106} (\bibinfo {year}
  {2007})\BibitemShut{NoStop}%
\bibitem{Qian2006}%
  \BibitemOpen
  \bibfield{author}{%
  \bibinfo {author} {\bibfnamefont{X.}~\bibnamefont{Qian}}, \bibinfo {author}
  {\bibfnamefont{J.}~\bibnamefont{Li}}, \bibinfo {author}
  {\bibfnamefont{X.}~\bibnamefont{Lin}},\ and\ \bibinfo {author}
  {\bibfnamefont{S.}~\bibnamefont{Yip}},\ }%
  \bibfield{journal}{%
  \bibinfo {journal} {Phys. Rev. B}\ }%
  \textbf{\bibinfo {volume} {73}},\ \bibinfo {pages} {035408} (\bibinfo {year}
  {2006})\BibitemShut{NoStop}%
\bibitem{Perdew2003}%
  \BibitemOpen
  \bibfield{author}{%
  \bibinfo {author} {\bibfnamefont{J.}~\bibnamefont{Perdew}}\ and\ \bibinfo
  {author} {\bibfnamefont{S.}~\bibnamefont{Kurth}},\ }%
  in\ \emph{\bibinfo {booktitle} {A Primer in Density Functional Theory}},\
  \bibinfo {editor} {edited by\ \bibinfo {editor}
  {\bibfnamefont{C.}~\bibnamefont{Fiolhais}}, \bibinfo {editor}
  {\bibfnamefont{F.}~\bibnamefont{Nogueira}},\ and\ \bibinfo {editor}
  {\bibfnamefont{M.}~\bibnamefont{Marques}}}\ (\bibinfo {publisher}
  {Springer-Verlag},\ \bibinfo {year} {2003})\ p.~\bibinfo {pages}
  {9}\BibitemShut{NoStop}%
\bibitem{Doltsinis2002}%
  \BibitemOpen
  \bibfield{author}{%
  \bibinfo {author} {\bibfnamefont{N.~L.}\ \bibnamefont{Doltsinis}}\ and\
  \bibinfo {author} {\bibfnamefont{D.}~\bibnamefont{Marx}},\ }%
  \bibfield{journal}{%
  \bibinfo {journal} {J. Th. Comp. Chem.}\ }%
  \textbf{\bibinfo {volume} {1}},\ \bibinfo {pages} {319} (\bibinfo {year}
  {2002})\BibitemShut{NoStop}%
\bibitem{DiVentra2000}%
  \BibitemOpen
  \bibfield{author}{%
  \bibinfo {author} {\bibfnamefont{M.}~\bibnamefont{Di Ventra}}\ and\ \bibinfo
  {author} {\bibfnamefont{S.~T.}\ \bibnamefont{Pantelides}},\ }%
  \bibfield{journal}{%
  \bibinfo {journal} {Phys. Rev. B}\ }%
  \textbf{\bibinfo {volume} {61}},\ \bibinfo {pages} {16207} (\bibinfo {year}
  {2000})\BibitemShut{NoStop}%
\bibitem{Verlet1967}%
  \BibitemOpen
  \bibfield{author}{%
  \bibinfo {author} {\bibfnamefont{L.}~\bibnamefont{Verlet}},\ }%
  \bibfield{journal}{%
  \bibinfo {journal} {Phys. Rev.}\ }%
  \textbf{\bibinfo {volume} {159}},\ \bibinfo {pages} {98} (\bibinfo {year}
  {1967})\BibitemShut{NoStop}%
\bibitem{Castro2004}%
  \BibitemOpen
  \bibfield{author}{%
  \bibinfo {author} {\bibfnamefont{A.}~\bibnamefont{Castro}}, \bibinfo {author}
  {\bibfnamefont{M.~A.~L.}\ \bibnamefont{Marques}},\ and\ \bibinfo {author}
  {\bibfnamefont{A.}~\bibnamefont{Rubio}},\ }%
  \bibfield{journal}{%
  \bibinfo {journal} {J. Chem. Phys.}\ }%
  \textbf{\bibinfo {volume} {121}},\ \bibinfo {pages} {3425} (\bibinfo {year}
  {2004})\BibitemShut{NoStop}%
\bibitem{Bao2003}%
  \BibitemOpen
  \bibfield{author}{%
  \bibinfo {author} {\bibfnamefont{W.}~\bibnamefont{Bao}}, \bibinfo {author}
  {\bibfnamefont{D.}~\bibnamefont{Jaksch}},\ and\ \bibinfo {author}
  {\bibfnamefont{P.~A.}\ \bibnamefont{Markovich}},\ }%
  \bibfield{journal}{%
  \bibinfo {journal} {J. Comp. Phys.}\ }%
  \textbf{\bibinfo {volume} {187}},\ \bibinfo {pages} {318} (\bibinfo {year}
  {2003})\BibitemShut{NoStop}%
\bibitem{Mortensen2005}%
  \BibitemOpen
  \bibfield{author}{%
  \bibinfo {author} {\bibfnamefont{J.~J.}\ \bibnamefont{Mortensen}}, \bibinfo
  {author} {\bibfnamefont{L.~B.}\ \bibnamefont{Hansen}},\ and\ \bibinfo
  {author} {\bibfnamefont{K.~W.}\ \bibnamefont{Jacobsen}},\ }%
  \bibfield{journal}{%
  \bibinfo {journal} {Phys. Rev. B}\ }%
  \textbf{\bibinfo {volume} {71}},\ \bibinfo {pages} {035109} (\bibinfo {year}
  {2005})\BibitemShut{NoStop}%
\bibitem{Enkovaara2010}%
  \BibitemOpen
  \bibfield{author}{%
  \bibinfo {author} {\bibfnamefont{J.}~\bibnamefont{Enkovaara}} \emph{et~al.},\
  }%
  \bibfield{journal}{%
  \bibinfo {journal} {J. Phys. Cond. Mat.}\ }%
  \textbf{\bibinfo {volume} {22}},\ \bibinfo {pages} {253202} (\bibinfo {year}
  {2010})\BibitemShut{NoStop}%
\bibitem{Perdew1992}%
  \BibitemOpen
  \bibfield{author}{%
  \bibinfo {author} {\bibfnamefont{J.}~\bibnamefont{Perdew}}\ and\ \bibinfo
  {author} {\bibfnamefont{Y.}~\bibnamefont{Wang}},\ }%
  \bibfield{journal}{%
  \bibinfo {journal} {Phys. Rev. B}\ }%
  \textbf{\bibinfo {volume} {46}},\ \bibinfo {pages} {12947} (\bibinfo {year}
  {1992})\BibitemShut{NoStop}%
\bibitem{Li2005}%
  \BibitemOpen
  \bibfield{author}{%
  \bibinfo {author} {\bibfnamefont{X.}~\bibnamefont{Li}}, \bibinfo {author}
  {\bibfnamefont{J.~C.}\ \bibnamefont{Tully}}, \bibinfo {author}
  {\bibfnamefont{H.~B.}\ \bibnamefont{Schlegel}},\ and\ \bibinfo {author}
  {\bibfnamefont{M.~J.}\ \bibnamefont{Frisch}},\ }%
  \bibfield{journal}{%
  \bibinfo {journal} {J. Chem. Phys.}\ }%
  \textbf{\bibinfo {volume} {123}},\ \bibinfo {pages} {084106} (\bibinfo {year}
  {2005})\BibitemShut{NoStop}%
\bibitem{Blum2009}%
  \BibitemOpen
  \bibfield{author}{%
  \bibinfo {author} {\bibfnamefont{V.}~\bibnamefont{Blum}}, \bibinfo {author}
  {\bibfnamefont{R.}~\bibnamefont{Gehrke}}, \bibinfo {author}
  {\bibfnamefont{F.}~\bibnamefont{Hanke}}, \bibinfo {author}
  {\bibfnamefont{P.}~\bibnamefont{Havu}}, \bibinfo {author}
  {\bibfnamefont{V.}~\bibnamefont{Havu}}, \bibinfo {author}
  {\bibfnamefont{X.}~\bibnamefont{Ren}}, \bibinfo {author}
  {\bibfnamefont{K.}~\bibnamefont{Reuter}},\ and\ \bibinfo {author}
  {\bibfnamefont{M.}~\bibnamefont{Scheffler}},\ }%
  \bibfield{journal}{%
  \bibinfo {journal} {Comp. Phys. Comm.}\ }%
  \textbf{\bibinfo {volume} {180}},\ \bibinfo {pages} {2175} (\bibinfo {year}
  {2009})\BibitemShut{NoStop}%
\end{thebibliography}

\end{document}